\documentclass[preprint]{iucr} 
  \journalcode{J}
  \papertype{CP} 
  
                \def\href#1{\relax}\let\foo\caption
\ifPDF
\RequirePackage{hyperref}
\PassOptionsToPackage{pdftex,bookmarksopen,bookmarksnumbered}{hyperref}
\fi
\let\caption\foo
\usepackage{amsmath}
\usepackage{graphicx}
\usepackage{tikz}
\usepackage{url}
\usepackage[mode=buildnew]{standalone}
\usetikzlibrary{calc,positioning,shapes,arrows,matrix,fit,plotmarks,patterns,patterns.meta,mindmap,decorations.markings,decorations.pathreplacing,decorations.pathmorphing,intersections,fadings,arrows.meta,bending,decorations.text,arrows.meta}
\tikzset{>=stealth}

\definecolor{Tradewind}{HTML}{66C2A5}
\definecolor{Salmon}{HTML}{FC8D62}
\definecolor{PoloBlue}{HTML}{8DA0CB}
\definecolor{Shocking}{HTML}{E78AC3}
\definecolor{Kimberly}{RGB}{0,121,52}

\newcommand\inputpgf[2]{{
		\let\pgfimageWithoutPath\pgfimage
		\renewcommand{\pgfimage}[2][]{\pgfimageWithoutPath[##1]{#1/##2}}
		\let\includegraphicsWithoutPath\includegraphics
		\renewcommand{\includegraphics}[2][]{\includegraphicsWithoutPath[##1]{#1/##2}}
		\input{#1/#2}
}}

\graphicspath{./figures/}

\begin{document}                  % DO NOT DELETE THIS LINE

     %-------------------------------------------------------------------------
     % The introductory (header) part of the paper
     %-------------------------------------------------------------------------

     % The title of the paper. Use \shorttitle to indicate an abbreviated title
     % for use in running heads (you will need to uncomment it).

\title{\textit{NeuXtalViz}: Interactive Three-Dimensional Visualization and Analysis for Single-Crystal Neutron Diffraction}
%\shorttitle{Short Title}

     % Authors' names and addresses. Use \cauthor for the main (contact) author.
     % Use \author for all other authors. Use \aff for authors' affiliations.
     % Use lower-case letters in square brackets to link authors to their
     % affiliations &  if there is only one affiliation address, remove the [a].

\cauthor[a]{Zachary}{Morgan}{morganzj@ornl.gov}{}

\author[a]{Zhongcan}{Xiao}

\author[a,b,c]{Sylwia}{Pawledzio}

\author[d]{Iris}{Ye}

\author[e]{Kathleen}{Loughlin}

\author[f]{Shiyun}{Jin}

\author[a]{Vickie}{Lynch}

\author[a]{Thomas}{Proffen}

\author[a]{Christina}{Hoffmann}%\email{iiauthor@gmail.com}
%\equalcont{These authors contributed equally to this work.}

\cauthor[a]{Xiaoping}{Wang}{wangx@ornl.gov}{}%\email{iiiauthor@gmail.com}
%\equalcont{These authors contributed equally to this work.}

\aff[a]{Neutron Scattering Division, Oak Ridge National Laboratory, One Bethel Valley Rd, Oak Ridge, 37831, TN, USA}

\aff[b]{Spanish CRG BM25-SpLine Beamline at European Synchrotron Radiation Facility, Grenoble 38000, France
}

\aff[c]{Instituto de Ciencia de Materiales de Madrid-ICMM/CSIC, Cantoblanco,
Madrid E-28049, Spain
}

\aff[d]{Next Generation STEM Internship Participant}

\aff[f]{Gemological Institute of America, 5355 Armada Drive, Carlsbad, 92008, CA, USA}

\aff[e]{Department of Materials Science \& Engineering, University of Tennessee, Knoxville, TN
37996, USA}

%\aff[a]{Neutron Scattering Division, Oak Ridge National Laboratory}
%\aff[b]{Next Generation Pathways to Computing at Oak Ridge National Laboratory Participant}

     % Use \shortauthor to indicate an abbreviated author list for use in
     % running heads (you will need to uncomment it).

%\shortauthor{Soape, Author and Doe}

     % Use \vita if required to give biographical details (for authors of
     % invited review papers only). Uncomment it.

%\vita{Author's biography}

     % Keywords (required for Journal of Synchrotron Radiation only)
     % Use the \keyword macro for each word or phrase, e.g. 
     % \keyword{X-ray diffraction}\keyword{muscle}

%\keyword{keyword}

     % PDB and NDB reference codes for structures referenced in the article and
     % deposited with the Protein Data Bank and Nucleic Acids Database (Acta
     % Crystallographica Section D). Repeat for each separate structure e.g
     % \PDBref[dethiobiotin synthetase]{1byi} \NDBref[d(G$_4$CGC$_4$)]{ad0002}

%\PDBref[optional name]{refcode}
%\NDBref[optional name]{refcode}

\maketitle                        % DO NOT DELETE THIS LINE

\begin{abstract}
\textit{NeuXtalViz} (Neutron Single-Crystal Visualization) is a Python-based software package developed at Oak Ridge National Laboratory to provide interactive three-dimensional visualization and analysis tools for single-crystal neutron diffraction experiments. Built on the \textit{Mantid} framework for data reduction, and leveraging \textit{PyVista} and \textit{Matplotlib} within a Python Qt environment, \textit{NeuXtalViz} adopts a model--view--presenter architecture that separates the user interface from the core processing components. The software provides a unified interface for tasks central to single-crystal diffraction, including $UB$-matrix determination, experiment planning, visualization of normalized reciprocal-space volumes, and real-space crystal-structure calculations. It also integrates with widely used community tools and has been deployed on instrument and analysis servers, where it is now being adopted by instrument teams and users. By embedding advanced three-dimensional visualization directly into the experimental workflow, \textit{NeuXtalViz} enhances the planning, execution, and analysis cycle for single-crystal neutron diffraction experiments, while providing a flexible framework for future development.
\end{abstract}

     %-------------------------------------------------------------------------
     % The main body of the paper
     %-------------------------------------------------------------------------
     % Now enter the text of the document in multiple \section's, \subsection's
     % and \subsubsection's as required.

\section{Introduction and background} \bigskip

Neutron diffraction is a powerful bulk probe for crystal-structure analysis, offering capabilities that complement X-ray diffraction, particularly for resolving nuclear structures containing both light and heavy elements and for determining complex magnetic order. When suitably sized crystals are available, single-crystal diffraction provides especially rich information because it preserves the underlying three-dimensional crystallographic information that is spherically averaged in powder diffraction.

Processing single-crystal diffraction data is intrinsically three-dimensional, requiring more complex data structures, algorithms, and visualization tools than are typically needed for one-dimensional diffraction patterns. Three-dimensional visualization therefore remains an important challenge: data must be displayed intuitively, manipulated with sufficient fidelity, and updated efficiently enough to support interactive analysis. Relevant 3D data types include reciprocal-space volumes, such as momentum-transfer or reciprocal-lattice-unit histograms containing diffraction peaks; real-space crystallographic unit cells with atomic coordinates; and raw diffraction data collected by large-area detectors over many crystal orientations, sometimes with varying incident wavelength \cite{mcintyre_area_2015}.

Oak Ridge National Laboratory (ORNL) hosts several dedicated single-crystal diffractometers, all with signature characteristic large, position sensitive area detectors \cite{coates_suite-level_2018,wang_single-crystal_2025}.
Among these includes the time-of-flight white beam diffractometers TOPAZ, MANDI, SNAP, and CORELLI at the Spallation Neutron Source (SNS) \cite{schultz_integration_2014,coates_macromolecular_2015,ye_implementation_2018,massani_single-crystal_2020}.
At the High Flux Isotope Reactor (HFIR), there are the monochromatic instruments DEMAND and WAND$^2$ and the soon-to-be rebuilt and upgraded quasi-Laue IMAGINE instrument \cite{cao_demand_2019,frontzek_wand2versatile_2018,meilleur_imagine_2020}.
In order to provide a robust user experience during beamtime experiments, it is desirable to have high-peformance, interactive 3D visualization tools that integrate the latest advances in data-reduction processing with a clear path toward incorporating state-of-the-art analysis programs. Previous efforts toward this goal have often resulted in separate tools that proved difficult to maintain over the long term. For example, the \textit{Integrated Software Analysis Workbench} (\textit{ISAW}), developed at the Intense Pulse Neutron Source (IPNS), provided a Java-based graphical user interface (GUI) with a ``3D-event'' viewer that displayed histograms of neutron event data from instruments such as TOPAZ using thresholding~\cite{mikkelson_coordinated_2002}. Similarly, the original \textit{MantidPlot} software included the \textit{VATES} (Visualization and Analysis Toolkit Extensions) viewer~\cite{arnold_mantiddata_2014}, which supported non-orthogonal crystallographic reciprocal-space histograms via ``3D-volume'' axes, built on top of the \textit{ParaView} framework~\cite{ahrens_paraview_2005} and exposing many of its visualization features. The current \textit{Mantid} interface, called \textit{Workbench}, no longer incorporates these capabilities, although it retains a 3D-to-2D slice viewer for volume histograms.

To address this need, a new software package called \textit{NeuXtalViz} (pronounced \textit{new-crystal-vis}, short for Neutron Single-Crystal Visualization) has been developed to enhance the user experience by embedding 3D visualization within an interactive interface. In its current form, it focuses on several core features: (1) $UB$-matrix determination, (2) experiment planning, (3) volumetric visualization of combined and normalized reciprocal-space data, and (4) crystal-structure calculations. In addition, it provides links to widely used analysis software, including an interface for structural refinements with \textit{SHELX}~\cite{sheldrick_short_2008} and diffuse-scattering simulations with \textit{DISCUS}~\cite{proffen_discus_1997}. It also offers plugins for reduction workflows and other utilities such as an experiment summary tool to the ORNL Neutron Catalog utilizing \textit{PyONCat} module.

\section{Program architecture} \bigskip

\begin{figure}
    \includegraphics[width=\linewidth]{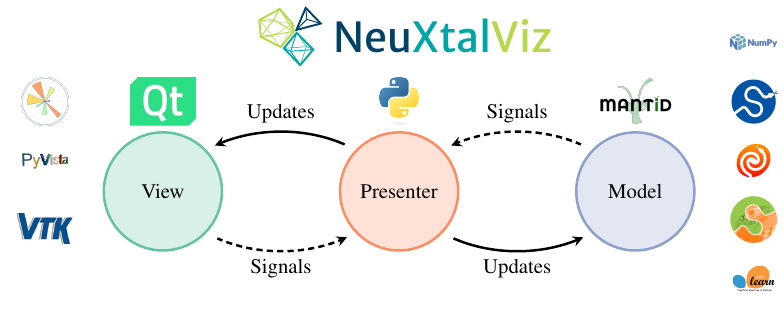}\\
    \caption{Overview of the model-view-presenter architecture and the corresponding libraries used in the \textit{Qt} view (e.g. \textit{Matplotlib} and \textit{PyVista}/\textit{VTK}) and \textit{Mantid} model (leveraging other libraries such as \textit{NumPy}, \textit{SciPy}, \textit{Astropy}, \textit{scikit-image} and \textit{scikit-learn}. The presentation layer is pure Python, handling all logic and connection between the view and presenter without explicitly knowing about the modules inside either the view or model.}
    \label{fig:mvp}
\end{figure}

\textit{NeuXtalViz} is a Python program that leverages Mantid \cite{arnold_mantiddata_2014} for behind-the-scenes processing already in use at the SNS and HFIR single-crystal diffractometers, while making heavy use of \textit{PyVista} \cite{sullivan_pyvista_2019} and Matplotlib \cite{hunter_matplotlib_2007} for visualization, all integrated within a Python Qt framework as summarized in Fig.~\ref{fig:mvp}. It adopts a model-view-presenter architecture, inspired by the Mantid software itself, to achieve two key goals. First, it enforces a clear separation of concerns: the interactive interface is well-isolated from the core processing machinery by a presentation layer. Although this design is not the most sophisticated and can introduce additional boilerplate code, it provides a practical, maintainable structure well-suited for scientific software developed at a national user facility. Second, it facilitates more effective testing, particularly at the model level for unit testing, while also offering a straightforward pathway to automate user documentation. 

One major achievement of this code layout is the provision of a common look and feel across all main applications. The central layout is built around a base view with a centrally located 3D visualization powered by \textit{PyVista}. This allows both real- and reciprocal-space data to be readily displayed, including histograms of reciprocal-space events, peak positions, sizes, intensities, atomic positions, and related quantities. A common set of buttons is provided to orient the view along specific crystallographic directions ($\boldsymbol{a}$, $\boldsymbol{b}$, $\boldsymbol{c}$, $\boldsymbol{a}^\ast$, $\boldsymbol{b}^\ast$, $\boldsymbol{c}^\ast$) or reciprocal-space Cartesian axes (beam $\pm Q_z$, vertical $\pm Q_y$, in-plane horizontal $\pm Q_x$) attached to the sample. The corresponding base model requires a sample orientation and crystal-to-Cartesian transformation (the $UB$ matrix \cite{busing_angle_1967} formalism). The base presenter manages all button connections to reorient or reset the 3D view, save plots, and display essential sample information such as lattice parameters and crystal orientation.

\section{Requirements and availability} \bigskip

The program is available at \url{https://github.com/neutrons/NeuXtalViz-tools}. An up-to-date installation is also maintained on the instrument and user analysis servers at ORNL, enabling seamless use both at the beamlines during experiments and on user machines before and after experiments for planning and analysis, respectively. A shortcut on the computers activates a \textit{conda} environment with the required libraries and launches the GUI within that environment. Access to experiment data is provided via a mounted file system on the analysis computers. 

A local installation is also possible, but is generally not recommended unless required for development. In that case, installation is achieved by cloning the \textit{GitHub} repository, creating the \textit{conda} environment from the provided list of software dependencies, and performing a \textit{pip} installation in editable mode. Mounting the instrument experiment data can be facilitated using a service such as \texttt{sshfs}. The GUI is launched through a simple \texttt{NeuXtalViz.py} script. A \textit{PyPI} and/or \textit{Conda} package may be made available in the future. Documentation is supported through \textit{Sphinx}, built using \textit{GitHub Actions} workflow. The docstrings follow \textit{numpydoc} style. A dedicated webpage is located at \url{https://neutrons.github.io/NeuXtalViz-tools/source/NeuXtalViz.html} with tutorials with for users.

\section{Applications and examples} \bigskip

Each primary application within \textit{NeuXtalViz} is accessible through a unified interface and can be enabled or switched on demand. The current suite includes \textit{UB}, \textit{Experiment Planner}, \textit{Volume Slicer}, and \textit{Crystal Structure}, with clear pathways for integrating additional tools in the future. These applications rely heavily on algorithms provided by \textit{Mantid}. Detailed descriptions of individual algorithms and low-level components are therefore omitted, as comprehensive documentation is available at \url{https://www.mantidproject.org/}. Supplementary technical details and workflow conventions specific to single-crystal diffraction are provided at \url{https://single-crystal.ornl.gov/}. In addition, \textit{NeuXtalViz} supports plugins to external software, including tools developed for ORNL single-crystal diffractometers as well as third-party applications accessed via command-line interfaces or direct links to installed analysis software on the user analysis server.

\subsection{UB Determination}

A key critical step of a single-crystal diffraction experiment is determining the $UB$-matrix that relates the crystal reciprocal lattice axes vector to laboratory reference frame through the goniometer rotation matrix $R$ \cite{busing_angle_1967}. It is a required step for experiment planning and reducing data via peak integration, reciprocal space reconstruction, and/or order parameter tracking. It relates the measured momentum transfer laboratory wave vector $\boldsymbol{Q}$ to the reciprocal lattice $(h,k,l)$ frame.

\begin{figure}
    \includegraphics[width=\linewidth]{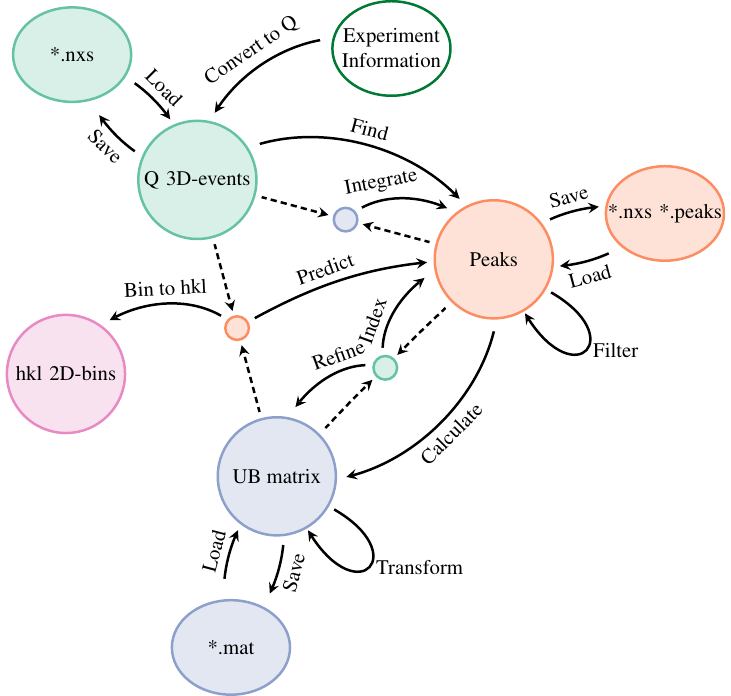}\\
    \caption{Schematic illustrating the key interactions involved in $UB$-matrix determination. Three \textit{Mantid} workspaces are defined: (1) a 3D reciprocal-space event workspace, (2) a peaks table, and (3) the sample unit-cell $UB$ matrix. Arrows denote data flow, with tails indicating the required input workspace(s) and heads indicating the resulting or updated workspace. Intermediate nodes indicate operations requiring multiple workspaces. Annotations identify specific sub-workflows or \textit{Mantid} algorithms.}
    \label{fig:ub}
\end{figure}

The determination of the $UB$ matrix for time-of-flight Laue diffractometers at ORNL has traditionally relied on beamline-specific scripts developed independently for each instrument. In addition, a limited graphical interface embedded in the former \textit{MantidPlot} environment provided a direct route to $UB$ determination through a basic sequence of data-reduction steps. This functionality was removed during the transition to \textit{MantidWorkbench} and was not replaced with an equivalent integrated graphical workflow. The original tool supported conversion of event data into reciprocal $Q$-sample space, identification of strong diffraction peaks, initial unit-cell determination, and application of the conventional orientation transform. The present implementation reproduces this core functionality while extending it to support beamline-specific workflows and providing meaningful, interactive visualization of the data at each stage of the process.

Rather than implementing each instrument workflow independently, the strength of the MVP architecture is leveraged by first defining a clear and minimal model that specifies the core objects and their interactions. This design is summarized in Fig.~\ref{fig:ub}, which illustrates the principal components of the model. The first object is a three-dimensional reciprocal-space dataset generated from user-specified experimental inputs, including Integrated Proposal Tracking System (IPTS) and run numbers, together with basic instrument metadata such as the instrument name, wavelength settings, and optional calibration files. The second object is a peaks table containing information on harvested diffraction peaks, including their positions in reciprocal space, indexing, and integrated intensities. The third object is an oriented reciprocal lattice, which encapsulates the unit-cell parameters and their uncertainties together with the crystal orientation matrix $U$.

From the lattice parameters, the corresponding $B$ matrix is constructed via a Cholesky decomposition of the reciprocal-lattice metric tensor. The crystal orientation is commonly represented by the vectors $\boldsymbol{u}$, $\boldsymbol{v}$, and $\boldsymbol{w}$, which define the reciprocal-lattice directions in Miller $(h,k,l)$ space aligned with the incident beam, the horizontal direction perpendicular to the beam, and the vertical direction, respectively. These directions are defined in the reference configuration where all goniometer axes, both fixed and movable, are conceptually returned to their home positions (even if this configuration is not physically realizable), such that the rotation matrix reduces to the identity, $R = I$.

\subsubsection{TOPAZ-strategy}

\begin{figure}
    \includegraphics[width=\linewidth]{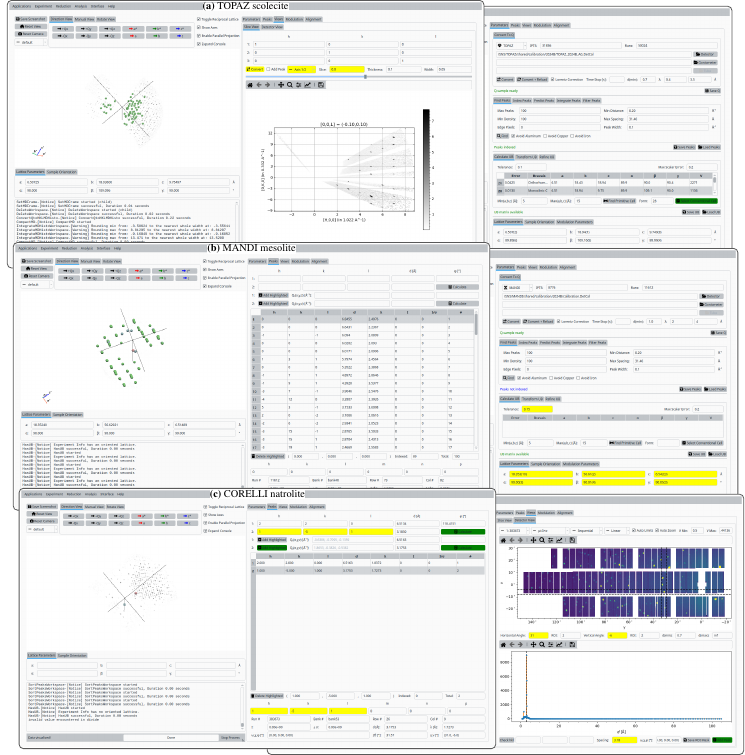}\\
    \caption{$UB$ determination in \textit{NeuXtalViz}. Selected screenshots illustrating three strategies for Laue diffraction data: (a) TOPAZ scolecite data, (b) MANDI mesolite data, and (c) CORELLI natrolite data. In (a), strong reflections are harvested from reciprocal-space volumes to determine an initial primitive (Niggli-reduced) unit cell, which is subsequently transformed to the conventional base-centered monoclinic setting. In (b), strong peaks are likewise identified, but an initial unit-cell guess for the face-centered cell is provided a priori and refined under orthorhombic constraints. In (c), two reflections are manually selected and indexed, and the crystal orientation is calculated using known lattice parameters. In all cases, the orientation and lattice parameters are refined by predicting additional reflections, refining peak positions and optimizing the $UB$ matrix.}
    \label{fig:ubnxv}
\end{figure}

A straightforward and robust approach for assessing crystal quality and determining the unit cell is to harvest a large number of strong diffraction peaks and identify the underlying periodicity in reciprocal space. In this method, an initial primitive unit cell is obtained by searching for reciprocal-lattice basis vectors, where the distribution of peaks projected along candidate directions is evaluated using Fourier analysis \cite{steller_algorithm_1997}. This procedure yields a reduced primitive unit cell, which can subsequently be transformed into one of several compatible conventional unit cells drawn from a set of crystallographically allowed candidates \cite{mighell_lattice_2001}. An illustration of this approach is shown in Fig.~\ref{fig:ubnxv}(a) for scolecite ($\mathrm{CaAl_2Si_2O_{10}\cdot3H_2O}$), which crystallizes in a pseudo-tetragonal monoclinic $Cc$ structure \cite{kvick_neutron_1985}.

A key strength of this approach is that it requires only minimal prior information about the approximate size of the primitive unit cell, which is inherently constrained by the observed diffraction data. As a result, both the unit cell and the corresponding $UB$ matrix can be refined without prior knowledge of the crystal structure. This capability is particularly valuable for rapid crystal screening and quality assessment, providing an independent consistency check in chemical crystallography experiments. Moreover, it facilitates the identification of overlooked symmetries or alternative candidate unit cells and Bravais lattices, especially when complemented by the sensitivity of neutron diffraction to light elements such as hydrogen.

\onecolumn
%\twocolumn

Once an initial unit cell has been determined, it can be refined by constraining the lattice parameters according to the assumed crystal symmetry. A rapid projection of the data into reciprocal space using user-specified crystallographic directions enables efficient screening of reciprocal-space slices, providing a qualitative assessment of the $UB$ refinement based on how well diffraction peaks align with integer grid axes. This visualization also facilitates the identification of fractional or satellite reflections. The refined $UB$ matrix is then saved for subsequent use.

\begin{figure}
    \includegraphics[width=\linewidth]{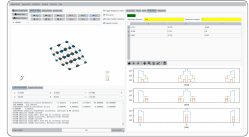}\\
    \caption{Modulation analysis tools in \textit{NeuXtalViz}. Satellite reflections are folded into a reciprocal-lattice extended $2 \times 2 \times 2$ cell containing the first Brillouin zone. Reflection clusters are identified using DBSCAN, with user-defined parameters for the minimum neighbor distance in reciprocal lattice units and the minimum number of peaks required to define a cluster. The centroid of each cluster is reported in a table as a candidate modulation vector, together with histograms of the offsets in $h$, $k$, and $l$.}
    \label{fig:modulation}
\end{figure}

In cases where fractional satellite peaks are present, a dedicated modulation utility is available as shown in Fig.~\ref{fig:modulation}. Given the refined $UB$ matrix, all peaks are folded back into the first Brillouin zone. Density-based clustering using the DBSCAN algorithm \cite{ester_density-based_1996}, as implemented in \textit{scikit-learn} \cite{pedregosa_scikit-learn_2011}, is then applied to identify candidate modulation vectors. User-defined parameters, including the minimum cluster size and cutoff distance, are used to detect clusters whose centroids correspond to potential modulation wave vectors. In this implementation, care is taken to correctly identify the principal (main-lattice) cluster while accounting for the centrosymmetry of the neutron diffraction pattern. The resulting modulation vectors may then be used to define modulation offsets \cite{smaalen_elementary_2004}, including higher-order harmonics and cross terms as supported within \textit{Mantid}. Optionally, these modulation vectors may be further refined by optimizing a modulated $UB$ matrix, which is linked to the primary $UB$ matrix.

\subsubsection{MANDI-strategy}

In structural biology, protein crystals are characterized by substantially larger unit cells and are most commonly described by a limited subset of lower-symmetry space groups \cite{mighell_lattice_2001}, reflecting their organic and macromolecular nature. Neutron diffraction is primarily employed in this context to locate hydrogen and deuterium atoms, and as a result, samples delivered to the beamline are typically already well-characterized using X-ray diffraction. 

Within this framework, strong diffraction peaks are first identified in the same manner as for inorganic crystals. However, rather than performing a reduced-cell search, the known lattice parameters may be supplied directly, allowing the expected unit cell to be solved. The resulting $UB$ matrix can then be refined using the measured peak positions and saved for subsequent experiment planning and data reduction. An example of this approach is shown in Fig.~\ref{fig:ubnxv}(b) for mesolite ($\mathrm{Na_2Ca_2Si_9Al_6O_{30}\cdot8H_2O}$), another zeolite from the natrolite family, which crystallizes in space group $Fdd2$ and features a pronounced long axis of approximately $56\,\text{\AA}$ \cite{artioli_x-ray_1986}.

\subsubsection{CORELLI/SNAP-strategy}

In many condensed-matter physics experiments, the crystallographic unit cell is already well established. When probing phenomena such as magnetism, experiments are often planned to exploit high-symmetry directions of the crystal, preferentially aligning the sample to access specific crystallographic planes or directions. Moreover, complex sample environments—including cryomagnets, cryostats, pressure cells, and furnaces—can significantly restrict detector coverage and available sample rotations. As a result, crystals are commonly pre-aligned prior to data collection.

Adopting an approach analogous to that used in spectroscopy experiments, the crystal may be incrementally rotated while searching for strong Bragg reflections. The resulting datasets can be loaded into the application, where peaks are selected manually using a \textit{matplotlib}-based instrument view that displays neutron counts mapped onto spherical horizontal and vertical angles. The data may additionally be cropped by $d$-spacing range, and a region of interest surrounding a candidate reflection can be examined as a one-dimensional intensity profile in  $d$-spacing. Once identified, the peak may be manually added to the peaks table.

With a minimum of two non-collinear reflections, the peaks table utility supports manual indexing of selected peaks. Given known lattice parameters, the $UB$ matrix can then be computed directly by fixing the lattice constants and refining only the crystal orientation. Using this initial $UB$ matrix, additional reflections are predicted in reciprocal space down to a specified resolution cutoff, subject to the appropriate lattice-centering condition. The predicted peak positions may subsequently be refined by automatically shifting each reflection to its local center of mass within a defined cutoff radius in reciprocal space. This constrained refinement procedure can be iterated to further improve the accuracy of the $UB$ matrix while optionally refining or constraining the known lattice parameters. This workflow is illustrated in Fig.~\ref{fig:ubnxv}(c) for natrolite ($\mathrm{Na_2Al_2Si_3O_{10}\cdot2H_2O}$), which crystallizes in the space group $Fdd2$ \cite{artioli_neutron_1984,ross_crystalline_1992}.

\subsubsection{WAND$^2$/DEMAND-strategy}

A $UB$ matrix may be determined from first principles using the tool by applying any of the procedures described above for monochromatic instruments. In practice, however, the $UB$ matrix is more commonly determined directly by the instrument control software \cite{lumsden_ub_2005,lumsden_spicespectrometer_2006}, which is particularly well suited to monochromatic diffraction where orientation calculations can be performed without invoking the full data-reduction workflow. 

The typical use case therefore involves loading the experimental data, extracting the initial $UB$ matrix embedded within the metadata, and subsequently predicting and refining diffraction peak positions. In some cases, the volume slicing utility can be used to assess indexing quality by visually inspecting the alignment of peaks with reciprocal-lattice directions. Additionally, the tool supports transformation of an existing $UB$ matrix by supplying a positive-definite matrix that linearly maps the original indexing basis into a new reference frame.

\onecolumn
\begin{figure}
    \includegraphics[width=\linewidth]{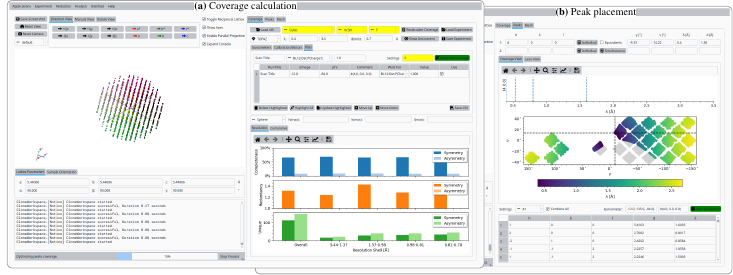}\\
    \caption{Experiment planning in \textit{NeuXtalViz} for silicon at TOPAZ. (a) Coverage calculation after adding a selected sample orientation to the experiment plan. The 3D view shows predicted peaks to a specified resolution cutoff, symmetrized using the cubic $m\bar{3}m$ point group. Overall and resolution-shell completeness, redundancy, and the number of unique reflections are calculated and plotted with and without symmetry. (b) Peak placement tool showing the possible detector locations of a specified reflection on the TOPAZ instrument view, colored by wavelength.
}
    \label{fig:planner}
\end{figure}
%%\twocolumn

\subsection{Experiment Planner}

Determination of the $UB$ matrix is only the first step in the initial data-reduction workflow and serves as the foundation for subsequent experiment planning. Experiment planning primarily involves selecting the goniometer angles to measure, given the available rotational degrees of freedom and the instrument’s reciprocal-space coverage, which is defined by its solid-angle detector geometry and accessible wavelength range(s).

The \textit{CrystalPlan} software historically provided core capabilities for optimizing reciprocal-space coverage and targeting specific reflections for several diffractometers at ORNL, with particularly heavy use on the TOPAZ beamline \cite{zikovsky_crystalplan_2011}. However, the original implementation relied on now-deprecated versions of Python and \textit{wxPython}, did not interface with \textit{Mantid}, and employed \textit{Mayavi} for three-dimensional visualization \cite{ramachandran_mayavi_2011}. While \textit{Mayavi} offers functionality similar to \textit{PyVista}, it does not provide a straightforward mechanism for direct integration with \textit{VTK} in a manner well suited to crystallographic visualization, particularly when preserving aspect ratios, angular relationships, and lattice geometry without distortion.

The present tool re-implements much of the core functionality of \textit{CrystalPlan} while introducing substantial improvements with improved performance. In particular, the transition to a \textit{Mantid}-based workspace model enables straightforward extension to instruments already defined within the \textit{Mantid} repository. Furthermore, functionality refined through years of \textit{CrystalPlan} usage by the TOPAZ instrument team is combined with scripts developed by the CORELLI team, resulting in a comprehensive and unified set of experiment-planning utilities.

As in \textit{CrystalPlan}, the $UB$-matrix file is loaded, an instrument and goniometer configuration are selected, and optional calibration files may be specified. The desired resolution is then defined together with optional crystal-symmetry information, including the point group and lattice centering. Expected counting statistics, based on either acquisition time or accumulated proton charge, can be evaluated for each orientation. The resulting experiment plan can be saved as a NeXus file for later reuse or exported as a comma-separated values (CSV) file for execution as EPICS table scans. Figure~\ref{fig:experiment}(a) shows a screenshot of the application with the calculated coverage accounting for crystal symmetry.

\subsubsection{Coverage optimizer}

The \textit{CrystalPlan} algorithm has been re-implemented using \textit{Mantid} instrument definition files and workspaces to enable automated experiment planning. The method employs evolutionary optimization to select crystal orientations that efficiently sample reciprocal space, reducing manual trial-and-error in single-crystal neutron diffraction experiments.

Rather than manually selecting goniometer angles, the algorithm generates many candidate orientation sets, predicts which Bragg reflections would be measurable for each set, and evaluates how completely these orientations sample reciprocal space within a chosen resolution and wavelength range accounting for crystal symmetry and previously chosen angles. Candidate plans are then ranked by coverage, and the most effective ones are retained and recombined to produce improved plans in subsequent iterations.

Through repeated cycles of selection, recombination, and random variation, the algorithm progressively converges on a small set of orientations that maximize symmetry-aware reflection completeness while respecting instrument and experimental constraints. The final result is an optimized measurement plan that improves data completeness and reduces redundant measurements, enabling more efficient use of beam time.

This version of the \textit{CrystalPlan} algorithm achieves efficiency comparable to the original implementation. However, improved reciprocal-space coverage is obtained by preferentially weighting lower-resolution shells during optimization, which enhances overall completeness by emphasizing low-$Q$ regions that are typically more difficult to sample due to diffraction geometry.

\subsubsection{Peak locator}

In many single-crystal diffraction experiments, it is necessary to identify a single goniometer setting at which a specific Bragg reflection is accessible within the instrument’s coverage, while avoiding detector edges or wavelength band cutoffs. As in its predecessor, the experiment planner provides the capability to simulate reciprocal-space coverage for an individual reflection. This functionality is extended by visualizing the coverage of the selected reflection across the entire instrument, represented by a colormap indicating the contributing wavelength, and by explicitly displaying any harmonic reflections [Fig.~\ref{fig:planner}(b)]. Symmetry-equivalent reflections may also be included based on the selected crystallographic point group.

A further enhancement allows a second reflection to be specified for simultaneous accessibility with the first. This capability, introduced in \textit{NeuXtalViz}, is implemented by identifying the intersection of goniometer angle settings for which both reflections are simultaneously accessible. This feature is particularly useful for instruments such as TOPAZ, whose ambient goniometer configuration provides multiple rotational degrees of freedom. Future work will extend this approach to monochromatic beamlines, where accessibility will be computed directly from scattering angles rather than wavelength bands, which are more natural for Laue diffractometers.

A selected orientation can be added directly to the current experiment plan. In addition, a combined ``Laue'' view is provided that displays all Bragg reflections predicted for a given orientation, together with a table listing the corresponding indexed peaks and their calculated $d$-spacings and wavelengths $\lambda$.

\subsubsection{Coverage simulator}

Support is also provided for simulating reciprocal-space slices with a variety of projections, analogous to those available in the \textit{UB} tool. Instead of visualizing measured data, these views simulate instrument coverage based either on the currently planned goniometer angles or on a specified mesh scan, defined by the number of orientations sampled for each angle within user-selected minimum and maximum limits. The latter approach is generally preferred for instruments with a single rotational degree of freedom and is inspired by the direct-geometry spectroscopy planning tool \textit{DGSPlanner} embedded within \textit{MantidWorkbench}. 

For instruments equipped with multiple goniometer motors, the additional degrees of freedom make it possible to define a mesh scan directly from two target $(h,k,l)$ directions. Given a desired effective rotational range and step size, the goniometer settings are optimized such that the selected reflection directions remain predominantly within the equatorial scattering plane throughout the scan. On instruments such as TOPAZ equipped with an ambient goniometer, this typically produces a scan strategy in which the innermost rotation axis first aligns one reflection direction into the scattering plane, while the outermost axis performs the primary scan motion with smaller compensating adjustments from the inner axis to maintain the desired geometry. Such scan trajectories are particularly advantageous for diffuse scattering measurements, where continuous and efficient coverage of reciprocal space is important for reconstructing three-dimensional pair distribution function maps.

\begin{figure}
    \includegraphics[width=\linewidth]{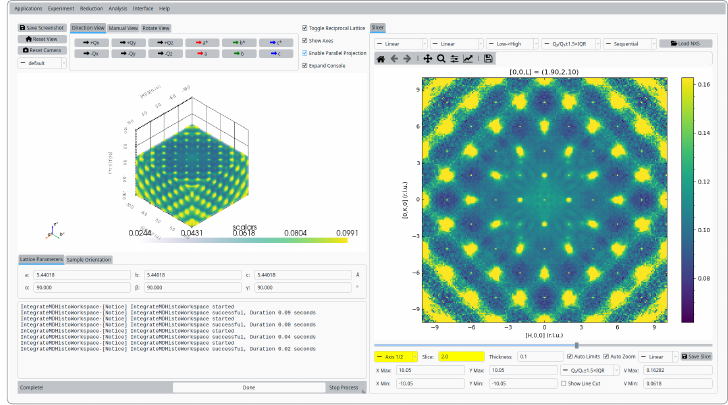}\\
    \caption{Volume slicer in \textit{NeuXtalViz}. A normalized reciprocal-space volume of silicon measured on TOPAZ is shown, highlighting thermal diffuse scattering around Bragg peaks. The displayed slice corresponds to the $(hk2)$ scattering plane. Interactive controls are provided for integrating slices, adjusting color scales, and zooming within the volume.}
    \label{fig:volume}
\end{figure}

\subsection{Reciprocal Space Volume Slicer}

One of the key capabilities of modern single-crystal diffraction software suites is the mapping and combination of reciprocal-space data into a normalized three-dimensional volume for both structure determination and diffuse scattering analysis \cite{michels-clark_expanding_2016,savici_efficient_2022}. Visualization of such data in three dimensions can reveal salient features, including modulated order and short-range correlations. These volumes are typically expressed in reciprocal lattice units through linear combinations of the reciprocal lattice vectors. Preserving the correct aspect ratios and interaxial angles is therefore essential, particularly for non-orthogonal crystal systems such as hexagonal and monoclinic lattices.

To address this need, a dedicated volume slicer tool has been developed to provide accurate renderings of volumetric data with support for both two-dimensional slices and one-dimensional cuts. While plotting three-dimensional data in Cartesian coordinates with orthonormal basis vectors is straightforward using \textit{VTK} \cite{schroeder_visualizing_2000} and uniform grids (e.g., via \textit{ParaView}), support for non-orthogonal coordinate systems in \textit{PyVista} is achieved by applying an explicit transformation matrix that skews and stretches the axes to reproduce the correct lattice geometry.

From these voxelized grids, users can interactively explore the volume using the \textit{PyVista} clipping and slicing widgets [Fig.~\ref{fig:volume}]. Selecting a slice axis and position, or dynamically sweeping through the volume, reveals two-dimensional slabs of reciprocal-space intensity. The extracted slice data may also be replotted using \textit{matplotlib}, where curvilinear axes are used to preserve the underlying affine transformation, enabling interactive zooming and direct indexing in reciprocal lattice coordinates $(h,k,l)$. In addition, one-dimensional rods or cuts may be generated along either in-plane axis of the slice to reveal corresponding intensity profiles.

Although currently limited to axis-aligned slicing and cuts, the volume slicer does not require the data to be expressed strictly along the conventional $[h,0,0]$, $[0,k,0]$, and $[0,0,l]$ directions. Arbitrary in-plane linear combinations of reciprocal-lattice vectors may instead be used to define the slicing axes through an appropriate affine transformation. For example, a common projection for hexagonal crystal systems employs the $[h,h,0]$ and $[-k,k,0]$ directions, which are orthogonal in reciprocal space and separated by $90^\circ$. This capability allows natural visualization of high-symmetry planes and directions without resampling or rebinning the underlying volumetric data, while still preserving correct metric relationships and indexing.

In addition to standard options for displaying volumetric data using a variety of colormaps and intensity scalings (e.g., linear and logarithmic), several new features have been introduced to automatically scale the data based on statistical properties of the voxel intensities. Rather than relying on simple minimum and maximum values, which are often dominated by strong Bragg reflections, outlier-based scaling methods are employed to enhance features that rise above the background yet remain weaker than the strongest peaks.

One such approach uses $\mathrm{MEAN} \pm 3\,\mathrm{STD}$, where $\mathrm{STD}$ denotes the standard deviation of the intensity distribution. A more robust alternative is given by $\mathrm{MED} \pm 1.5\,\mathrm{MAD}$, where $\mathrm{MED}$ and $\mathrm{MAD}$ are the median and median absolute deviation, respectively \cite{weng_k-space_2020}. Finally, a method based on Tukey’s criterion is implemented using $Q_{1,3} \pm 1.5\,\mathrm{IQR}$, where $\mathrm{IQR}$ is the interquartile range and $Q_{1}$ and $Q_{3}$ denote the first and third quartiles, respectively \cite{morgan_rmc-discord_2021}.

\subsection{Crystal Structure Factor Calculator}

\begin{figure}
    \includegraphics[width=\linewidth]{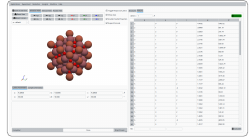}\\
    \caption{Crystal structure factor calculator in \textit{NeuXtalViz}. The refined structure of the mineral bixbyite, including site mixing between iron and manganese, is shown. Squared structure-factor amplitudes are calculated and displayed in tabular form.}
    \label{fig:crystal}
\end{figure}

A dedicated tool for working with crystal structures is also available. Figure~\ref{fig:crystal} displays a selected screenshot visualizing a crystal structure. Originally developed within the \textit{rmc-discord} framework to leverage \textit{PyVista} for the modification, generation, and visualization of unit cells and supercells \cite{morgan_rmc-discord_2021}, this functionality has since been adapted for integration into \textit{NeuXtalViz}. The tool supports direct import of crystallographic information files (CIFs) \cite{brown_cif_2002}, which are rendered as atomic arrangements within the unit cell. Individual atomic sites may be interactively edited, including updates to fractional coordinates, occupancies, and atomic species. For neutron diffraction applications, specific isotopes may be assigned, with scattering lengths updated accordingly.

Through integration with \textit{Mantid}, the tool can compute nuclear structure factors either to a user-defined resolution cutoff or for symmetry-equivalent reflections, with the results presented in tabulated form. Modified crystal structures may then be exported as \textit{SHELX} \texttt{.ins} files for subsequent structure refinement \cite{sheldrick_short_2008}.

Although intentionally lightweight, this utility can also support $UB$ determination by providing reference information for identifying characteristic reflections and their associated $d$-spacings. Future extensions could incorporate intensity simulations using virtual instrument configurations, further bridging structure definition and experiment planning.

\subsection{Additional interfaces}

Many crystallographic analysis packages remain driven by command-line workflows and text-based configuration files. To bridge this gap, \textit{NeuXtalViz} provides a lightweight interface that gives users direct access to these installed tools without leaving the graphical environment. Users can navigate to a working directory, open and edit input files, and execute external programs directly from within the application.

This interface currently supports widely used refinement tools from the \textit{SHELX} suite, including \texttt{shelxl} and \texttt{shelxt}, as well as the \textit{DISCUS} \texttt{discus\_suite}. By lowering the barrier between interactive visualization and script-based analysis, this functionality opens the door to more integrated workflows, including future use cases that combine conventional structure refinement with diffuse-scattering simulations.

\begin{figure}
    \includegraphics[width=\linewidth]{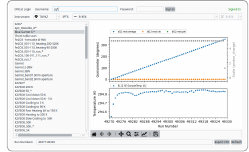}\\
    \caption{Experiment browser in \textit{NeuXtalViz}. Metadata for a selected experiment (rare-earth garnet collected at TOPAZ) is queried using ORNL Neutron Catalog (ONCat) services \cite{parker_oncat_2018} and summarized by run title, providing an organized overview of the associated measurement runs.}
    \label{fig:experiment}
\end{figure}

Additional utilities include an experimental data browser that provides a concise, graphical overview of single-crystal runs used during data reduction and analysis [Fig.~\ref{fig:experiment}]. Acting as a digital logbook, this tool organizes runs by title and generates condensed, reduction-ready identifiers, while simultaneously visualizing goniometer angle settings counting statistics, and temperature for each run. By presenting this information in a compact and intuitive format, the browser offers immediate insight into experimental coverage and progression, streamlining data lookup and reducing the likelihood of transcription errors during analysis.

Finally, interfaces are provided to two complementary data-reduction tools. The first is a dedicated graphical interface for TOPAZ peak integration that supports parallel processing and enables users to read, write, and edit configuration files \cite{schultz_integration_2014}. The second is a closely related interface, \texttt{garnet}, which applies a similar parallel-processing framework to reciprocal-space normalization \cite{michels-clark_expanding_2016}. 

In addition to conventional reduction workflows, \texttt{garnet} supports event-based data filtering \cite{granroth_event-based_2018,fancher_time_2018}, allowing neutron events to be filtered by sample log parameters such as temperature, electric field, or magnetic field. The resulting data are represented as four-dimensional workspaces, comprising a three-dimensional normalized reciprocal-space volume augmented by an additional filtered dimension.

\subsection{External connections}

Beyond its internal toolset, \textit{NeuXtalViz} provides direct links to widely used, community-developed crystallographic software available on the facility Linux analysis servers. These include \textit{Olex2} \cite{dolomanov_olex2_2009} and \textit{ShelXle} \cite{hubschle_shelxle_2011}, which are commonly used for small-molecule structure solution and refinement through powerful graphical interfaces. Direct links to \textit{VESTA} are also provided, enabling high-quality visualization of crystal structures, electron-density maps, and magnetic moments \cite{momma_vesta_2011}.

Interfaces are additionally provided to the \textit{FullProf} suite, which is frequently employed for magnetic structure modeling and refinement \cite{rodriguez-carvajal_magnetic_2025}. Looking ahead, integration with \textit{Jana2020} would be a particularly valuable addition, offering advanced capabilities for the refinement of magnetic and modulated structures within a unified crystallographic framework \cite{petricek_jana2020_2023}.

\section{Conclusion and future} \bigskip

\textit{NeuXtalViz} has been developed at ORNL, driven by instrument teams and made available to neutron scattering staff since 2024, with partial evaluation and feedback from early adopters beginning in the 2025A SNS run cycle. Wider adoption by instrument teams followed during the 2025B run cycle, including the full replacement of the original \textit{CrystalPlan} software, which is no longer maintained in its original form.

The software provides essential interactive tools, leveraging both 3D and 2D visualization to help users more clearly observe the effects of parameter choices during $UB$ refinement on peak finding and projection selection, while representing crystallographic reciprocal-space data more naturally in chosen projections. These capabilities are presented in a consistent layout with a common look and feel across different applications.

Future work will focus on extending support for monochromatic diffraction workflows as DEMAND transitions from SPICE to EPICS. Additional opportunities include enabling end-to-end data reduction within a single environment, thereby minimizing the need to switch between external tools. Further development may also provide mechanisms for generating automated data-reduction configurations through the SNS web-based monitoring service, \url{https://monitor.sns.gov/dasmon/}. 

Incorporating live data visualization represents another promising direction, with the potential to provide immediate feedback during data collection and experiment planning \cite{bruhwiler_rapid_2021,kilpatrick_interactive_2023,kilpatrickmatthew_3d_2024}. Establishing a direct connection to instrument control systems remains a longer-term development goal, but one with potentially significant impact on experimental efficiency and user interaction \cite{xiao_crystalpilot_2025}. In addition, integration with advanced high-performance analysis workflows offers further opportunities to support scalable, near-real-time data processing and analysis \cite{yin_integrated_2024}. These advances position the framework as a scalable pathway toward autonomous, intelligent neutron experiments that accelerate scientific discovery.

Efforts are underway to integrate the software into broader initiatives aimed at developing a unified web-based and client interface for neutron scattering software \cite{watson_calvera_2022}. As these efforts continue to evolve, the present model--view--presenter (MVP) framework provides a practical and effective solution that already offers clear value to instrument scientists, experienced users, and new neutron diffraction users alike.

\section{Acknowledgments} \bigskip

Research sponsored by the Laboratory Directed Research and Development Program of Oak Ridge National Laboratory, managed by UT-Battelle, LLC, for the U.S. Department of Energy.
This research used resources at the High Flux Isotope Reactor and Spallation Neutron Source, a DOE Office of Science User Facility operated by the Oak Ridge National Laboratory. 
The beam time was allocated to TOPAZ, CORELLI, and MANDI on proposal numbers IPTS-31856, 31429, 34720, respectively.
This manuscript has been authored by UT-Battelle, LLC, under Contract No.
DE-AC0500OR22725 with the U.S. Department of Energy. The United States
Government retains and the publisher, by accepting the article for publication,
acknowledges that the United States Government retains a non-exclusive, paid-up,
irrevocable, world-wide license to publish or reproduce the published form of this
manuscript, or allow others to do so, for the United States Government purposes.
The Department of Energy will provide public access to these results of federally
sponsored research in accordance with the DOE Public Access Plan (http://energy.gov/
downloads/doe-public-access-plan).

\bibliographystyle{iucr}

%\bibliography{main.bi}

%\reference{Author, A. \& Author, B. (1984). \emph{Journal} textbf{Vol},  first page--last page.}
%\end{references}

\end{document}